\documentclass[dvips]{article}

\usepackage{amssymb,latexsym}
\usepackage{amsmath}
\usepackage{graphicx}
\usepackage{array}
\usepackage{changebar}
\usepackage{vmargin}

\setmarginsrb{1in}{.5in}{1in}{1.in}{14pt}{0.3in}{14pt}{0.494in}

\begin{document}
\title{Role of solvent for globular proteins in solution}
\author{Andrey Shiryayev, Daniel. L. Pagan, James. D. Gunton
      \\ Department of Physics, Lehigh University, Bethlehem,
      P.A., USA
       18015\\
       D. S. Rhen\\
       Department of Materials Science,\\
        University of Cambridge, \\
         Cambridge, CB2 3QZ, UK\\
        Avadh Saxena and Turab Lookman,\\
       Theoretical Divsion, Los Alamos National Laboratory,\\
       Los Alamos, NM 87545 USA}
\date{\today}
\maketitle

\begin{abstract}
The properties of the solvent affect the behavior of the solution.
We propose a model that accounts for the contribution of the
solvent free energy to the free energy of globular proteins in
solution. For the case of an attractive square well potential, we
obtain an exact mapping of the phase diagram of this model without
solvent to the model that includes the solute-solvent
contribution. In particular we find for appropriate choices of
parameters  upper critical points, lower critical points and even
closed loops with both upper and lower critical points, similar to
one found before \cite{Moelbert_03_01}. In the general case of
systems whose interactions are not attractive square wells, this
mapping procedure can be a first approximation to understand the
phase diagram in the presence of solvent.  We also present
simulation results for both the square well model and a modified
Lennard-Jones model.
\end{abstract}

\maketitle

\section{Introduction}

In recent years there has been an enormous increase in the number
of proteins that can be isolated, due to the rapid advances in
biotechnology. However, the determination of the function of these
proteins has been slowed by the difficulty of determining their
crystal structure by standard X-ray crystallography.  A major
problem is that it is difficult to grow good quality protein
crystals.  Experiments have clearly shown that this
crystallization depends sensitively on the physical factors of the
initial solution of proteins. An important observation was made by
George and Wilson \cite{Wilson_94_01}, who showed that x-ray
quality globular protein crystals only result when the second
virial coefficient, $B_2$, of the osmotic pressure of the protein
in solution lies within a narrow range. This corresponds to a
rather narrow temperature window. For large positive $B_2$,
crystallization does not occur on observable time scales, whereas
for large negative $B_2$, amorphous precipitation occurs.
Rosenbaum, Zamora and Zukoski then showed \cite{Zukoski_96_01}
that crystallization of globular proteins could be explained as
arising from attractive interactions whose range is small compared
with the molecule's diameter (corresponding to the narrow window
of $B_2$). In this case the gas-fluid coexistence curve is in a
metastable region below the liquidus-solidus coexistence lines,
terminating in a metastable critical point.

When a system undergoes a phase transition the change in the Gibbs
free energy $\Delta G$ consists of two terms - an enthalpy change
$\Delta H$ and and an entropy change $\Delta S$, with $\Delta
G=\Delta H-T\Delta S$. The change in the Gibbs free energy must be
negative in order for the transition to occur. The enthalpy change
is negative because in the separated state the more dense phase
has a larger number of contacts and therefore its contact energy
is lower. However, the dense state has a lower entropy and its
entropy change is also negative. These two terms compete and at
low temperatures the free energy change due to particles going
from the dilute phase to the dense phase is negative and  phase
separation occurs. However, above some temperature the entropy
loss exceeds the enthalpy loss, so that the change in the Gibbs
free energy is positive and phase separation does not occur.

 The solvent can change this picture
dramatically \cite{Vekilov_02_01}. Consider the free energy change
of the particle going from the dilute phase to the dense phase.
\begin{equation}
\Delta G = \Delta H_{solute} - T \Delta S_{solute} + \Delta
H_{solvent} - T\Delta S_{solvent} \label{EnergyChangeSeparation}
\end{equation}
Here $\Delta H_{solute}$ is the enthalpy change of the particle
between the dense and dilute phases, $\Delta S_{solute}$ is the
entropy of the solute particle change, $\Delta H_{solvent}$ is the
enthalpy change of the solvent and $\Delta S_{solvent}$ is the
entropy change of the solvent. The first two terms on the right
side of (\ref{EnergyChangeSeparation}) behave exactly as described
above. The last two terms, however can be either negative or
positive.

As an example of this, consider water. Water has a large number of
strong intermolecular hydrogen bonds in the bulk state. When one
adds a hydrophobic particle to water, some of these bonds break.
This leads to an increase in the entropy of water and to a
decrease in its enthalpy. However, the further rearrangement of
the water molecules around the solute particle leads to even
stronger bonds, thereby decreasing solvent entropy and decreasing
the solvent enthalpy. When particles aggregate the water molecules
return to the original bulk state so the entropy and the enthalpy
of the water increase to their original
values\cite{SolventExample_1, SolventExample_2, SolventExample_3,
SolventExample_4}. This is an example how the $\Delta S_{solvent}$
and the $\Delta H_{solvent}$ terms in
(\ref{EnergyChangeSeparation}) can be positive. If the sign of the
total enthalpy and the total entropy change is positive, then
decreasing temperature can lead to the opposite behavior, when the
system tends to be separated above some temperature and uniform
below. This leads to a lower critical point. This phenomenon has
been experimentally observed in different systems \cite{LCST_01,
LCST_02, LCST_03, LCST_04}. The most interesting example of a
lower critical point in context of our work is the phase diagram
of  sickle hemoglobin, which is thought to have an upside down
fluid fluid coexistence curve \cite{Palma_91_00}, as inferred from
the observation of an upside down spinodal curve.

Below the lower critical point the water molecules build strong
hydrogen bonds around solute particles preventing them from
aggregating. As one increases the temperature the free energy
decrease of the aggregating particles surpasses the free energy
gain of the water and the system  phase separates. If one
increases the temperature further, the total entropy loss due to
aggregation becomes larger than the total enthalpy loss, so the
phase separation becomes less favorable again and above some upper
critical temperature the system becomes uniform. Thus for the
hydrophobic particles one can have both an upper and a lower
critical point; in other words, the fluid-fluid phase diagram has
the form of a \textit{closed loop}. This can be the case for
different protein-water solutions.

\begin{figure}
\center
\includegraphics[width=10cm]{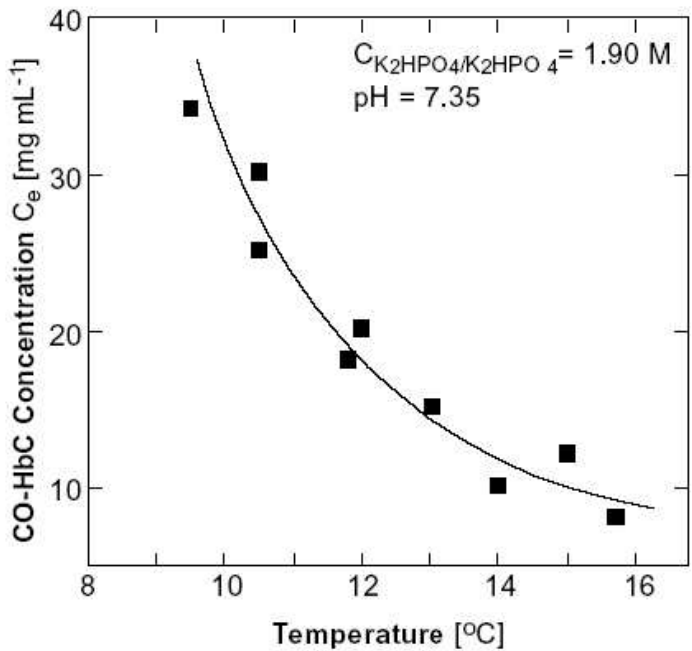}
 \caption{Solubility curve of CO HbC at conditions indicated at the plot.
 Points are the experimental results and the curve is fit using
 $\Delta H = 155 kJ/mol$ Details of the fitting procedure can be found in
 \cite{Vekilov_02_01}. The figure was nicely granted by P. G. Vekilov. }
 \label{fig_HbCSolubility}
\end{figure}

In this paper we present a simple model for the role of the
solvent.  We model the  multicomponent protein-solvent system  as
a binary system in which the solute molecule is much bigger than
the solvent. We assume that the role of other components in the
system are subsumed in the effective solute-solute interaction.
The role of a salt, for example, is assumed simply to screen the
charge on the proteins, inducing the effective inter particle
attraction, whose strength is controlled by the salt
concentration. This assumption permits us to predict three
qualitatively different effects of the role of the solvent for
globular proteins in solution, all of which are consistent with
experimental observation \cite{Vekilov_02_01, Christopher_98_01}.
The first is a negative enthalpy and entropy of crystallization,
which gives a normal liquid-solid line. Lysozyme is an example of
this, as it has a negative enthalpy of crystallization, of the
order of minus 75 $kJ^{-1}$.  The second is a positive enthalpy of
crystallization, which gives the upside down (retrograde) behavior
of coexistence curves. This has been seen in hemoglobin C
(CO-HbC), which has a high positive enthalpy of 155 $kJ^{-1}$ of
crystallization. This system exhibits a strong retrograde
solubility dependence on temperature (figure
\ref{fig_HbCSolubility}). Another example of retrograde behavior
is chymotrypsinogen A \cite{Lu_02_00}. The third is a zero
enthalpy of crystallization, which gives vertical coexistence
curves. Apoferritin has an enthalpy of crystallization close to
zero and has a solubility which is independent of temperature
(vertical behavior in the temperature-solubility diagram).

We consider a situation in which the solute particle-particle
interactions are described by a repulsive hard core together with
an attractive interaction. For example, the interparticle
potential might be described by the DLVO potential
\cite{Pellicane_03_01}, which consists of a Debye-H$\ddot{u}$ckel
interaction and a van der Waals interaction. Other model
potentials are the square well and modified Lennard-Jones models
that have been used successfully in recent years to describe
globular proteins in solution. If the attractive part is
short-ranged then these effective potentials can qualitatively
describe the canonical phase diagram of the globular protein
solution. We can therefore consider that solvent contribution was
accounted in these effective potentials and there is no need to
add the solvent-solute interactions into potential. Examples of
proteins that have canonical phase diagram are Lysozyme
\cite{Vekilov_03_03,Lu_02_00} and $\gamma D$-crystallin
\cite{Berland_92_01}. However for proteins that have completely
different phase diagram (like apoferritin or HbS, HbC) these
effective potentials fail to describe the phase diagram even
qualitatively. In this case the solvent contribution wasn't
successfully included into effective potential. The model
presented in this paper is an attempt to account for the solvent
effect to describe the phase diagrams of the proteins like
apoferritin, HbS, HbC at least qualitatively.

We then include a simplified solute-solvent interaction to the
solute-solute interaction, which is similar to one used recently
to describe hydrophobic interactions \cite{Moelbert_03_01}. Our
particular interest is globular proteins in solution, but our
treatment in principle includes other systems.

  The outline of the paper is as follows.
Section 2 contains a description of our model, while section 3
contains a discussion of the particular case of a square well
system with solvent.  For this particular choice of solute-solute
interaction, we are able to obtain the phase diagram of the
square-well potential model \textbf{with} solvent contribution
from the phase diagram of the square-well potential
\textbf{without} solvent contribution. Section 4 discusses three
cases of this mapping, for different choices of parameter values,
that yield upper critical points, lower critical points, and
closed loops, respectively. Section 5 contains a derivation of a
modified Clausius-Clapeyron equation in the presence of the
solvent, that takes into account the temperature dependence of the
effective interaction potential.  Section 6 contains the results
of a simulation of a modified Lennard-Jones model plus solvent.
The modified Lennard-Jones model was originally introduced to
describe the phase diagram and nucleation rates for globular
proteins in solution \cite{Frenkel_97_00}. We show that for a
particular choice of parameter values the solvent produces a
fluid-fluid coexistence curve with a lower critical point.  The
liquid-solid coexistence lines for this model are also shown.
Finally, section 7 contains a brief conclusion.

\section{Simple model for solvent-solute interaction}

In this section we will derive a phenomenological model that
describes the influence of the solvent on a system of interacting
protein particles. Consider a system of protein particles
interacting via a short ranged pairwise potential
$U(\vec{r}_{ij})$. The total energy of such a system is
\begin{equation}
U_0(\textbf{r}^N) = \frac{1}{2}\sum_{i \neq j}U_0(|\textbf{r}_i -
\textbf{r}_j|) \label{total_energy0}
\end{equation}
where the subscript $0$ denotes the potential energy in the
absence of the solvent. The corresponding Helmholtz free energy
$F_0(N,V,T)$ of the system is
\begin{equation}
F_0(N,V,T) = -k_BT \ln\left( \frac{1}{\Lambda^{3N}N!} \int
d\textbf{r}^N \exp(-\beta U_0(\textbf{r}^N)\right)
\label{free_energy0}
\end{equation}
where $N$ is the number of particles in a volume V, $\Lambda$ is
the thermal de Broglie wavelength, $\textbf{r}$ the coordinates of
the particles and $k_B$ is a Boltzman constant. To take into
account the effect of the solvent on this system, we note that a
particular solvent molecule can be either in the bulk or in a
"shell" region around a protein particle (see Fig. \ref{fig_MLG}).
We define the energy and entropy (in units of $k_B$) differences
between these two states to be $\varepsilon_w$ and $\Delta s_w$,
respectively. In general these quantities depend on the
temperature \cite{Braun_03_01, Moelbert_03_01}, i.e. $\epsilon_w =
\epsilon_w(T)$, $\Delta s_w = \Delta s_w(T)$. Thus the free energy
difference between a solvent molecule in the bulk and in the shell
is $\Delta f = \epsilon_w - k_BT\Delta s_w$. The total energy of
the protein-water system therefore is
\begin{equation}
U = U_0 + \sum_{i=1}^N (\epsilon_w - k_BT\Delta s_w)
n_w^{(i)}(\textbf{r}^N) \label{total_energy}
\end{equation}
where $n_w^{(i)}$ is the number of water molecules around the
$i^{th}$ particle. We now assume that this number is proportional
to the available area on the protein molecule surface and
approximate this area as  proportional to the maximum possible
number of contacts of the protein molecule with other protein
molecules. That is, the available energy is proportional to $n_c$
minus the actual number of contacts $n_p^{(i)}$.
\begin{equation}
n_w^{(i)}(\textbf{r}^N) \propto n_c - n_p^{(i)}(\textbf{r}^N)
\label{water_contacts}
\end{equation}
We absorb the proportionality coefficient into $\epsilon_w$ and
$\Delta s_w$. Next, define two particles to be in contact when the
distance between their centers of mass is less than some value
$\lambda_c\sigma$ and greater than $\sigma$, where $\sigma$ is the
hard core diameter. Define the function $\gamma(|\textbf{r}_i -
\textbf{r}_j|)$ to be equal to one if the distance between
particles $i$ and $j$ is less than $\lambda_c\sigma$, but greater
than $\sigma$ and is zero otherwise. If the distance between the
particles is less than their hard core diameter, the total energy
of the system is infinite, which is the value of
$U_0(|\textbf{r}_i - \textbf{r}_j| < \sigma)$, so one can define
$\gamma(|\textbf{r}_i - \textbf{r}_j| < \sigma)=0$ at that
distance. The number of contacts for the $i^{th}$ particle is then
given by
\begin{equation}
n_p^{(i)} = \sum_{i \neq j} \gamma(|\textbf{r}_i - \textbf{r}_j|)
\label{protein_contacts}
\end{equation}
Combining equations (\ref{total_energy}), (\ref{water_contacts})
and (\ref{protein_contacts}), we find that the total energy of the
system is
\begin{equation}
U(\textbf{r}^N) = \frac{1}{2}\sum_{i \neq j}U_0(|\textbf{r}_i -
\textbf{r}_j|) + (\epsilon_w - k_BT\Delta s_w)\left( Nn_c -
\sum_{i \neq j}\gamma(|\textbf{r}_i - \textbf{r}_j|) \right)
\label{total_energy1}
\end{equation}
We can combine the first and the last terms on the right hand side
and define a modified intermolecular interaction parameter
\begin{equation}
U(|\textbf{r}_i - \textbf{r}_j|) = U_0(|\textbf{r}_i -
\textbf{r}_j|) - 2(\epsilon_w - k_BT\Delta
s_w)\gamma(|\textbf{r}_i - \textbf{r}_j|) \label{modified_energy}
\end{equation}

In summary, our model assumes that a water molecule can be either
in the bulk or close to the protein particle (shell); the number
of water molecules near the particle is proportional to the
available area on the surface of the particle; the available area
on the protein surface is proportional to the maximum possible
number of protein-protein contacts minus the actual number of
contacts.

\section{Square well system with solvent}

\subsection{Implementation of the general theory}

In order to simplify this expression for the energy, we consider a
system with a repulsive hard core and an attractive square-well
(SW) potential, whose range of attraction is $\lambda_c$. We can
write $U_0$ in the form of $U_0 = -1/2 \sum_{i=1}^N \epsilon_0
n_p^{(i)}$, where $\epsilon_0$ is the well depth. The total energy
of such a system is
\begin{equation}
U(\textbf{r}^N) = -\frac{1}{2}\sum_{i=1}^N(\epsilon_0 +
2\epsilon_w - 2k_BT\Delta s_w)n_p^{(i)} + (\epsilon_w - k_BT\Delta
s_w)Nn_c \label{total_energySW}
\end{equation}
The first term on the right hand side is just the energy of the
square-well potential with the well depth $\tilde{\epsilon} =
\epsilon + 2\epsilon_w - 2k_BT\Delta s_w$. This simplification
allows us to map the phase diagram of this square-well potential
model  \textbf{with} solvent contribution onto the phase diagram
of the square-well potential \textbf{without} solvent
contribution. Denote the free energy of the square-well system
with well depth $\epsilon$ as $F_0(N,V,T; \epsilon)$. The second
term on the right hand side of (\ref{total_energySW}) does not
depend on the position of the particles. Therefore the free energy
of the SW system \textbf{with} solvent
\begin{equation}
F(N,V,T) = F_0(N,V,T; \tilde{\epsilon}) + (\epsilon_w - k_BT\Delta
s_w)Nn_c \label{free_energySW}
\end{equation}
The free energy density of the system is
\begin{equation}
f(\rho,T) = f_0(\rho, T; \tilde{\epsilon}) + \rho n_c(\epsilon_w -
k_BT\Delta s_w) \label{free_energy_density}
\end{equation}
where $f_0$ is the free energy density of the square-well system
(without solvent). Denote $\mu_0$ and $\omega_0$ as the chemical
potential and grand canonical free energy density respectively. In
this case $\mu_0 = \partial f_0 /
\partial\rho$ and $\omega_0 = f_0 - \mu_0\rho$. To obtain the
phase diagram for the square-well system one needs to solve the
phase coexistence conditions $\mu_0(\rho_1,T) = \mu_0(\rho_2,T)$
and $\omega_0(\rho_1,T) = \omega_0(\rho_2,T)$. Suppose that we
have obtained this phase diagram in some way (numerically, for
example). The square-well coexistence curve has the following
functional form
\begin{equation}
\frac{kT_{coex}}{\epsilon} = \tau(\rho) \label{coeqistence}
\end{equation}

To obtain the phase diagram of a square well system \textbf{with}
the solvent we should write the chemical potential and grand
canonical free energy in terms of $\mu_0$ and $\omega_0$.
\begin{eqnarray}
&\mu(\rho,T) = \mu_0(\rho,T; \tilde{\epsilon}) + n_c(\epsilon_w -
k_BT\Delta s_w) \notag \\
&\omega(\rho,T) = \omega_0(\rho,T; \tilde{\epsilon})
\label{coex_conditions}
\end{eqnarray}
As we can see the phase diagram for the protein solvent system
maps to the square well phase diagram with the well depth
$\epsilon_0 + 2\epsilon_w - 2kT\Delta s_w$:
\begin{equation}
\frac{kT_{coex}}{(\epsilon_0 + 2\epsilon_w - 2kT_{coex}\Delta
s_w)} = \tau(\rho) \label{coexistence1}
\end{equation}
Assuming that $\epsilon_w$ and $\Delta s_w$ do not depend on
temperature, the relationship between the phase diagrams for these
two models is given by
\begin{equation} k_BT_{coex} =
\frac{(\epsilon_0 + 2\epsilon_w)}{(1 + 2\Delta s_w \tau(\rho))}
\tau(\rho) \label{coexistence2}
\end{equation}

In summary, we are able to obtain the phase diagram for an
interacting protein system that includes the effect of solvent in
terms of the interacting protein system without solvent, assuming
that the protein-water attractive interactions are given by a
square well potential; that the range of the protein-water
interaction is equal to the range of the protein-protein
interaction $\lambda = \lambda_c$ and that the parameters
$\epsilon_w$ and $\Delta s_w$ are constants.

\section{Upper and lower critical points and closed loops}
\subsection{Upper and lower critical points}
In this section we consider examples of the phase diagram mapping
for the square well model with solvent. Let $\tau(\rho)$ in
(\ref{coeqistence}) be a fluid-fluid coexistence curve. The
critical point can be determined by the following condition
\begin{equation}
\frac{\partial kT_{coex}}{\partial\rho_{coex}} = 0
\label{critical_0}
\end{equation}
If we define the left hand side of (\ref{coexistence1}) as $G(kT)$
then
\begin{equation}
\partial kT_{coex}/\partial\rho_{coex} = \frac{d\tau(\rho)/d\rho}
{dG(kT)/dkT} \notag
\end{equation}
This means that the critical density of the modified system is the
same as the critical density of the original square well system.
The second derivative shows whether there is an upper or lower
critical point. Taking into account that $d\tau/d\rho = 0$ at the
critical point, we obtain
\begin{equation}
\frac{\partial^2kT}{\partial\rho^2} =
\frac{d^2\tau/d\rho^2}{dG/dkT}_{crit} \label{second_derivative}
\end{equation}
The original square well model has an upper critical point, so
$d^2\tau(\rho_{cr})/d\rho^2 < 0$. Therefore the sign of the
denominator determines the type of the critical point. If
$dG(kT_{cr})/dkT > 0$ then there is an upper critical point.

In our simplified model with constant $\epsilon_w$ and $\Delta
s_w$ we can calculate $dG/dkT$ explicitly $dG/dkT = (\epsilon_0 +
2\epsilon_w)/(\tilde{\epsilon}^2)$. This gives the first condition
for upper or lower critical points. The temperature in
(\ref{coexistence2}) should be positive. This gives the second
condition required for having upper or lower critical points.
\begin{eqnarray}
\epsilon_w > -\epsilon_0/2  \textit{,  } \Delta s_w > \epsilon_0 /
2kT^0_{cr}
\textit{  - upper critical point} \label{upper_conditions} \\
\epsilon_w < -\epsilon_0/2 \textit{,  } \Delta s_w < \epsilon_0 /
2kT^0_{cr}\textit{  - lower critical point}
\label{lower_conditions}
\end{eqnarray}
where $T^0_{cr}$ is the critical temperature of the original
system without solvent.

\subsection{Closed loop phase diagram}
\begin{figure}
\center
\includegraphics[width=13cm]{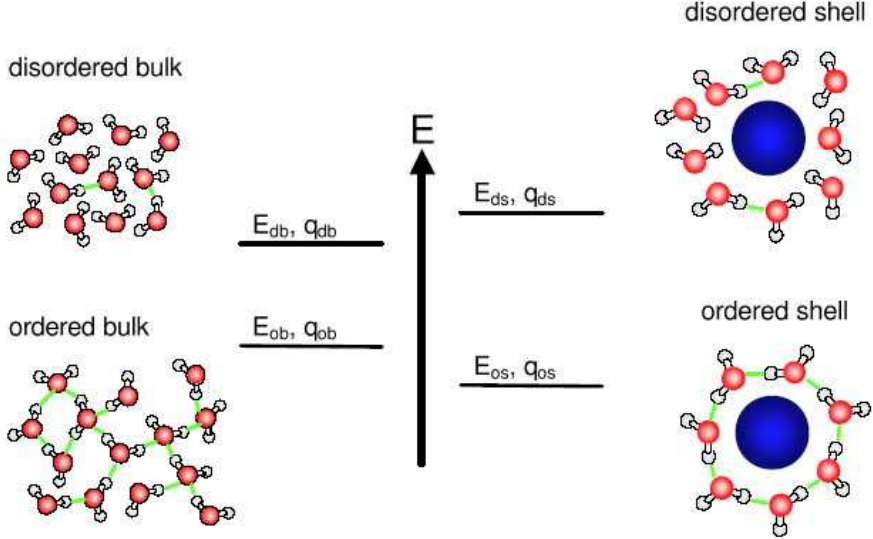}
 \caption{Energy levels for the MLG model for water. The
 configurations of the different states are shown schematically.
 The figure was nicely granted by S. Moelbert \cite{Moelbert_PhD}.
 } \label{fig_MLG}
\end{figure}

 One of the possible fluid-fluid coexistence curves
is a closed loop \cite{Moelbert_03_01} with both upper and lower
critical points. In order to obtain this behavior from our model
we must have temperature dependent $\epsilon_w$ and $\Delta s_w$.
We use the four level Muller, Lee, and Graziano (MLG)  model of
water \cite{Muller_90_00, Lee_96_00, Moelbert_03_01} to obtain
$\epsilon_w(kT)$ and $\Delta s_w(kT)$. In this model water
molecules can be either in disordered (broken hydrogen bonds) or
ordered states. When the protein particles are added, each state
of a water molecule splits into shell and bulk states. Thus we
have a four level model (Fig. \ref{fig_MLG}). The values of the
energy levels and degeneracies used in \cite{Moelbert_03_01} are
following : $E_{os} = -2$, $E_{ob} = -1$, $E_{db} = 1$, $E_{ds} =
1.8$, $q_{os} = 1$, $q_{ob} = 10$, $q_{db} = 40$ and $q_{ds} =
49$.  Moelbert and De Los Rios \cite{Moelbert_03_01} considered a
lattice model using the four level approximation and found by
Monte Carlo simulation a fluid-fluid coexistence line in the form
of a closed loop

\begin{figure}
\center
\includegraphics[width=12cm]{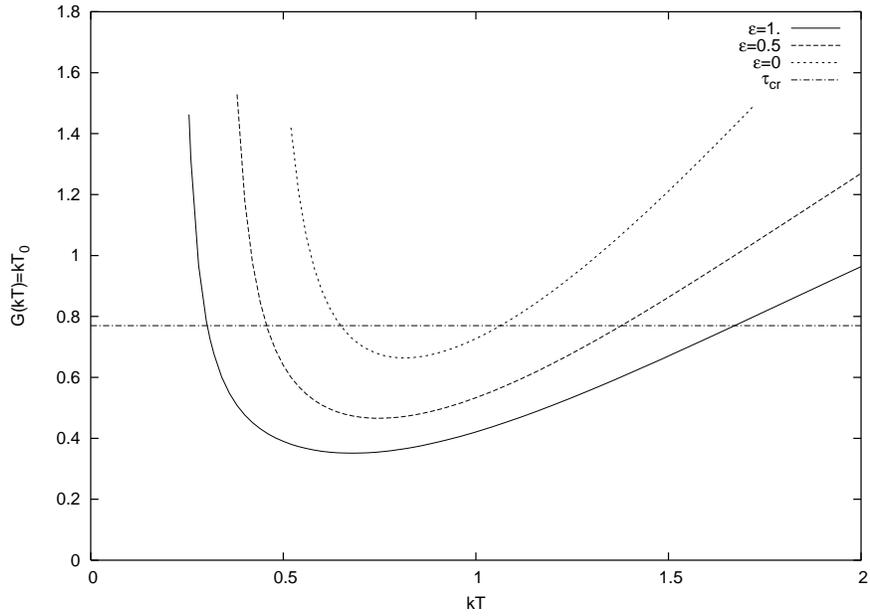}
 \caption{The dependence of the left hand side of (\ref{coexistence1})
 on the temperature for the MLG-type system (Mapping Curve) for 3 values of the protein-protein
 interaction energy. The horizontal dashed line shows
 the critical temperature of the original square well system (without solvent).
 All curves intersects\ the critical temperature of the system without
 solvent twice. This means that the MLG system in these cases has two
 critical points - one is the lower critical point (the slope of the curve is negative),
 the other is the upper critical point (the slope of the curve is positive).
 The phase diagram contains closed loop containing these two
 critical points.
  The parameters for the curves are the following $E_{os} = -2$,
 $E_{ob} = -1$, $E_{db} = 1$, $E_{ds} = 1.8$, $q_{os} = 1$, $q_{ob} = 10$,
 $q_{db} = 40$, $q_{ds} = 49$} \label{fig_loopMapping}
\end{figure}

Using the above values of energies and degeneracies we can
calculate the energy levels of the shell and bulk states:
\begin{eqnarray}
E_s = \frac{E_{os} + E_{ds}e^{-\beta\Delta E_s}}{1 +
e^{-\beta\Delta E_s}} \label{Energy_shell} \\
E_b = \frac{E_{ob} + E_{db}e^{-\beta\Delta E_b}}{1 +
e^{-\beta\Delta E_b}} \label{Energy_bulk}
\end{eqnarray}
In the same way we determine the values of the entropies for the
bulk and shell states:
\begin{eqnarray}
\frac{1}{k_B}S_s = ln( \frac{q_{os} + q_{ds}e^{-\beta\Delta
E_s}}{1 + e^{-\beta\Delta E_s}}) \label{Entropy_shell} \\
\frac{1}{k_B}S_b = ln( \frac{q_{ob} + q_{db}e^{-\beta\Delta
E_b}}{1 + e^{-\beta\Delta E_b}}) \label{Entropy_bulk}
\end{eqnarray}
where $\Delta E_s = E_{ds} - E_{os}$ and $\Delta E_b = E_{db} -
E_{ob}$. The parameters of our model $\epsilon_w$ and $\Delta s_w$
are the differences of the energies and entropies in the bulk and
shell states respectively,
\begin{eqnarray}
\epsilon_w(kT) = E_s - E_b \label{Energy_diff} \\
\Delta s_w(kT) = S_s - S_b \label{Entropy_diff}
\end{eqnarray}
Substituting (\ref{Energy_diff}) and (\ref{Entropy_diff}) into
(\ref{coexistence1}) we can numerically map the square well phase
diagram onto the MLJ-model phase diagram. Figure
\ref{fig_loopMapping} shows the dependence of the left hand side
of (\ref{coexistence1}) versus the temperature of the system
(mapping curve) in the presence of solvent for three different
values of the protein-protein interaction strength. The horizontal
line in fig. \ref{fig_loopMapping} represents the critical
temperature of the original square-well system without solvent.
The intersection of the mapping curve with the horizontal line is
the critical temperature of the system in the presence of solvent.
For each curve in fig. \ref{fig_loopMapping} there are two
intersections with the horizontal line. Therefore the mapping
procedure yields two critical points - one lower and one upper.The
mapping of the entire original coexistence curve gives a closed
loop type phase diagram for the MLG-type system.

 Using (\ref{Energy_shell}) -
(\ref{Entropy_bulk}) we can obtain the  low and high temperature
values of $\epsilon_w$ and $\Delta s_w$. This permits us to
determine the upper and lower critical points.  In the low
temperature limit
\begin{eqnarray}
\epsilon_w(0) = E_{os} - E_{ob} \label{low_kT} \\
\Delta s_w(0) = ln(q_{os}/q_{ob}) \notag
\end{eqnarray}
and in the high temperature limit:
\begin{eqnarray}
\epsilon_w(\infty) = \frac{E_{os} + E_{ds}}{2} -
\frac{E_{ob} + E_{db}}{2} \label{high_kT} \\
\Delta s_w(\infty) = ln(\frac{q_{os}+q_{ds}}{q_{ob}+q_{db}})
\notag
\end{eqnarray}
One necessary (but not sufficient) condition to have both upper
and lower critical points is (\ref{low_kT}) to satisfy conditions
(\ref{lower_conditions}) and (\ref{high_kT}) to satisfy
(\ref{upper_conditions}). The second condition would require that
the minimum of the mapping curve is lower than the critical
temperature of the original system without solvent (fig.
\ref{fig_loopMapping}).

\subsection{The case in which the range of the solute-solute interactions, $\lambda$,
differs from the solute-solvent interactions,  $\lambda_s$}

So far we have considered a special case in which the range of the
solute-solute interaction is equal to the that of the shell
region. In general this is not the case. In this section we
develop an approximate mapping for such systems. We use
Brilliantov's result \cite{Brilliantov_98_01} for the mean field
relations between the critical temperature and critical density:
\begin{equation}
\frac{kT_c}{v_0} = \rho_c \left[z_0+\frac{u_3}{2u_4}\right]_{CP} =
h(\rho_c) \label{BrilliantovRelationship}
\end{equation}
where $v_0$
\begin{equation}
v_0 = \int v(\textbf{r})d\textbf{r} \label{InteractionIntegral}
\end{equation}
is the zeroth order moment of the attractive part $v(\textbf{r})$
of the potential. The terms inside the brackets on the right hand
side of (\ref{BrilliantovRelationship}) are explained in
\cite{Brilliantov_98_01}. This  mean-field approximation has
different precision for different interactions and even for the
same interaction with different parameters. Division of the
potential into the repulsive and attractive parts adds more
freedom to obtaining parameter $v_0$ (\ref{InteractionIntegral})
\footnote{One of the methods to choose the form of the attractive
part of potential is to set pair-correlation function $g(r)$ equal
to zero for $r$ less than hard core diameter
\cite{Andersen_72_01}}, which also reduce quantitative precision.
The assumption made in this section is that the critical density
is the same for the systems with and without solvent. This is
approximately true for $\lambda_s < \lambda=1.25$, where
 the
critical density is about 0.4. For higher values of $\lambda$ this
assumption fails. The analysis given in this section therefore can
provide only qualitative understanding of the behavior of the
critical temperature for different $\lambda_s$.

What is important is that the right hand side of
(\ref{BrilliantovRelationship}) depends only on the isothermal
compressibility and its derivatives with respect to density of the
hard sphere system \cite{Brilliantov_98_01}. So for our purpose
this is just some function of the density and independent of
temperature.

The square well system in the absence of solvent has $v_0$ equal
to
\begin{equation}
v_0 = -\frac{4\pi\sigma^3}{3}(\lambda^3-1)\epsilon_0
\label{InteractionIntNoSolvent}
\end{equation}
where $\epsilon_0$ is the well depth of the particle particle
interaction. For the system in the presence of the solvent $v_0$
has the form:
\begin{equation}
v_0^s = -\frac{4\pi\sigma^3}{3}(\lambda^3-1)\epsilon_0 -
2(\epsilon_w - kT\Delta s_w)\frac{4\pi\sigma^3}{3}(\lambda_s^3-1)
= v_0\left( \frac{\epsilon_0}{\epsilon} + \frac{2(\epsilon_w -
kT_c^s\Delta s_w)}{\epsilon} \frac{\lambda_s^3-1}{\lambda^3-1}
\right) \label{InteractionIntNoSolvent}
\end{equation}
Substituting $v_0$ and $v_0^s$ into
(\ref{BrilliantovRelationship}) and assuming that the critical
density doesn't change we get the relationship between the
critical temperature of the system without solvent and the
critical temperature of the system with solvent:
\begin{equation}
kT_c^s = kT_c\left[ \frac{\epsilon_0}{\epsilon} +
\frac{2(\epsilon_w - kT_c^s\Delta s_w)}{\epsilon}
\frac{\lambda_s^3-1}{\lambda^3-1} \right]
\label{BrilliantovMappingCritical}
\end{equation}
Solving (\ref{BrilliantovMappingCritical}) for the $kT_c^s$ we get
the mapping relationship similar to (\ref{coexistence2}) but for
the critical temperature only:
\begin{equation}
k_BT_{coex} = \frac{(\epsilon_0 + 2\epsilon_w
\frac{\lambda_s^3-1}{\lambda^3-1})}{(1 + 2\Delta s_w \tau(\rho_c)
\frac{\lambda_s^3-1}{\lambda^3-1})} \tau(\rho_c)
\label{BrilliantovMappingCritical2}
\end{equation}
For the special case $\lambda_s$ equal to $\lambda$ this becomes
(\ref{coexistence2}) with $\tau_c = kT_c/\epsilon$. The main
difference is that the relationship (\ref{coexistence2}) is for
all temperatures, while the (\ref{BrilliantovMappingCritical2}) is
only for the critical point.  Let $\lambda_s^0$ be the value of
$\lambda_s$ at which the denominator on the right hand side of
(\ref{BrilliantovMappingCritical2}) vanishes. For $\lambda_s <
\lambda_s^0$ the system has an upper critical point behavior. For
$\lambda_s > \lambda_s^0$ the system has a lower critical point
behavior.

Using (\ref{BrilliantovMappingCritical}) we can approximately map
the phase diagram. In this case we can introduce the effective
solvent parameters
\begin{eqnarray}
\epsilon_w^{eff} &= \epsilon_w \frac{\lambda_s^3-1}{\lambda^3-1}
\notag \\
\Delta s_w^{eff} &= \Delta s_w \frac{\lambda_s^3-1}{\lambda^3-1}
\notag
\end{eqnarray}
These effective parameters correspond to the solute-solvent system
with an effective shell size $\lambda_s^{eff} = \lambda$,  with a
zeroth moment $v_0$ equal to $v_0^s$. So we approximately change
the system with solvent and $\lambda_s \neq \lambda$ by the system
with solvent and effective parameters $\epsilon_w^{eff}$ and
$\Delta s_w^{eff}$ with $\lambda_s^{eff} = \lambda$. This allows
us to perform a mapping of the SW system phase diagram onto the
effective system with solvent phase diagram using
(\ref{BrilliantovMappingCritical}). We can consider the latter as
the qualitative behavior of the system of interest.

\section{Modification of the Clausius-Clapeyron equation in the
presence of the solvent}

Now we consider a liquid-solid coexistence in the presence of the
solvent. To determine the solid-liquid coexistence curve we use
Gibbs-Duhem integration proposed by Kofke
\cite{Kofke_93_00,Kofke_93_01}. This method is based on the
integration of the Clausius-Clapeyron equation:
\begin{equation}
\frac{dP}{dT} = \frac{\Delta S}{\Delta V}
\label{Clausius_Clapeyron}
\end{equation}
where $\Delta S$ and $\Delta V$ are the entropy and volume
differences between solid and liquid phases, respectively. The
derivative is taken along the coexistence curve. In the absence of
the solvent this equation leads to the following form:
\begin{equation}
\frac{dP}{dT} = \frac{\Delta e + P\Delta v}{T\Delta v}
\label{Clausius_Clapeyron1}
\end{equation}
where $\Delta e = e_s - e_l$ and $\Delta v$ are the differences in
the energy and volume per particle in the solid and liquid phases
respectively.

When we have the model with a solvent contribution given by
(\ref{total_energy1}) or (\ref{total_energySW}), the interaction
potential is temperature dependent. Therefore one must modify the
(\ref{Clausius_Clapeyron1}) to take this temperature dependence
into account.  We start from the origins of the
(\ref{Clausius_Clapeyron}). For the coexistent phases the
difference of the Gibbs free energies is zero $\Delta G = G_s -
G_l = 0$. Thus the derivative is taken at the conditions of
$\Delta G=0$
\begin{equation}
\left(\frac{dP}{dT}\right)_{\Delta G=0} = -\frac{(\partial\Delta G
/\partial T)_{NP}}{(\partial\Delta G /\partial P)_{NT}} =
\frac{\Delta S}{\Delta V} \label{Clausius_Clapeyron2}
\end{equation}
To get $\Delta S$ we substitute the Gibbs partition function into
$S=-\partial G/\partial T$, where $G = -k_BT\ln\Xi(N,P,T)$, and
\begin{equation}
\Xi(N,P,T)= C\int dV d\textbf{r}^N
\exp[-\beta(U(\textbf{r}^N,T)+PV)]
\label{Gibbs_Partition_Function}
\end{equation}
The derivative of the Gibbs free energy with respect to
temperature then is
\begin{equation}
S = k\ln\Xi(N,P,T) + \frac{kT}{\Xi(N,P,T)} \int \left(
\frac{U+PV}{kT^2} - \frac{\partial U}{\partial T} \frac{1}{kT}
\right)e^{-\beta(U(\textbf{r}^N,T)+PV)} dV d\textbf{r}^N \notag
\end{equation}
which leads to
\begin{equation}
TS = -G + <U> + P<V> - T<\partial U/\partial T>
\label{Entropy_SolventContribution}
\end{equation}
Substituting this expression for the entropy into
(\ref{Clausius_Clapeyron}) we obtain the following modification of
the Clausius-Clapeyron equation
\begin{equation}
\frac{dP}{dT} = \frac{\Delta e + P\Delta v - T\Delta\partial
e}{T\Delta v}\label{Clausius_Clapeyron_New}
\end{equation}
where $\partial e = <\partial U/\partial T>/N$.

When the Hamiltonian of the system consists of two body
interactions between particles, the temperature is determined by
the average kinetic energy and a temperature dependent microscopic
Hamiltonian doesn't make much sense. In this case, however, we
have a system of solute particles in a solvent environment. A
rearrangement of the solute particles leads to a rearrangement of
solvent particles. The entropy of the system consists of two
parts: the entropy of the protein particles and the entropy of the
solvent molecules. Since in our model we don't consider separate
solvent molecules, but rather just average their contribution, the
effective interaction (\ref{modified_energy}) has the temperature
dependent term due to having integrated out the solvent degrees of
freedom.  This gives us the temperature dependent effective
interaction. The last term in equation
(\ref{Entropy_SolventContribution}) can be considered as the
solvent entropy contribution.

In the case of the square well protein-protein interactions, we
can use the potential (\ref{total_energySW}). The derivative of
this potential with respect to the temperature is
\begin{equation}
\frac{\partial U}{\partial T} = -\left(\frac{\partial
\epsilon_w}{\partial T} - k_BT \frac{\partial \Delta s_w}{\partial
T} - k_B\Delta s_w \right) \left( \sum_{i=1}^Nn_p^{(i)} - Nn_c
\right) \label{SW_PotentialDerivative}
\end{equation}
The $- Nn_c$ term in (\ref{SW_PotentialDerivative}) cancels when
we calculate the difference between solid and liquid phases. The
average of this derivative per particle gives $\partial e = \delta
(<n_p> - n_c)$, where $\delta$ is the expression in the first
parenthesis in (\ref{SW_PotentialDerivative}) and $<n_p>$ is the
average number of protein-protein contacts. Using
\ref{total_energySW} we can relate the average number of contacts
to the average energy. The difference between the average number
of contacts in the solid and  liquid phases is then related to the
$\Delta e$:
\begin{equation}
\Delta<n_p> = -\frac{2\Delta e}{\epsilon + 2\epsilon_w -
2k_BT\Delta s_w} \label{DeltaNumberContacts}
\end{equation}
The difference of $\partial e$ between two phases is therefore
\begin{equation}
\Delta\partial e = -\frac{2\delta\Delta e}{\epsilon + 2\epsilon_w
- 2k_BT\Delta s_w} \label{DeltaPartialE}
\end{equation}
Substituting (\ref{DeltaPartialE}) into
(\ref{Clausius_Clapeyron_New}) we obtain the final expression for
the Clausius-Clapeyron equation for the square-well model with
solvent
\begin{equation}
\frac{dP}{dT} = \left( \frac{\epsilon + 2\epsilon_w -
2\frac{\partial \epsilon_w}{\partial T} + 2k_BT \frac{\partial
\Delta s_w}{\partial T}} {\epsilon + 2\epsilon_w - 2k_BT\Delta
s_w} \Delta e + P\Delta v \right)\frac{1}{T\Delta
v}\label{Clausius_Clapeyron_SW}
\end{equation}

In the case of  a system in which the solvent-solute interactions
are not given by a square well potential, we have
\begin{equation}
\frac{dP}{dT} = \left( \Delta e + P\Delta v - 2\left[ \frac{
\frac{\partial \epsilon_w}{\partial T} - k_BT \frac{\partial
\Delta s_w}{\partial T} - k_B\Delta s_w}{\epsilon + 2\epsilon_w -
2k_BT\Delta s_w}\right] \Delta e^{SW} \right)\frac{1}{T\Delta
v}\label{Clausius_Clapeyron_SW}
\end{equation}

In equation (\ref{Clausius_Clapeyron1}) $\Delta e$, and $\Delta v$
are negative. The right hand side is therefore positive and the
coexistent pressure increases with increasing temperature. In
equation (\ref{Clausius_Clapeyron_SW}) $\Delta e$ and $\Delta v$
are again negative, but the term in parenthesis can be either
negative or positive; therefore for some cases the coexistent
pressure can decrease as we increase the temperature. This effect
is similar to the upside-down fluid-fluid coexistence curve and is
due to the entropy contribution from the solvent. While the
difference between the solid and the liquid of the entropy is
generally negative (the solid state is more ordered), the total
solvent plus solute entropy difference can be positive. This means
that the entropy increase of the solvent during  protein
crystallization due to the decrease of the contact number is
greater than the entropy decrease of the solute. Therefore the
numerator in (\ref{Clausius_Clapeyron}) can be positive for some
solvent parameters. At constant temperature $\Delta S = \Delta
H/T$. Thus this solvent model can reflect three different cases
experimentally observed by Vekilov \cite{Vekilov_02_01}.  One is a
negative enthalpy and entropy of crystallization, which gives the
normal liquid-solid line.  Another is a positive enthalpy of
crystallization, which gives the upside down behavior of
coexistence curves.  The third is a zero enthalpy of
crystallization, which gives the vertical coexistence curves.

\section{Numerical results for the solvent model}

\subsection{Summary of the mapping procedure and results for the
square well system}

\begin{figure}
\center
\includegraphics[width=12cm]{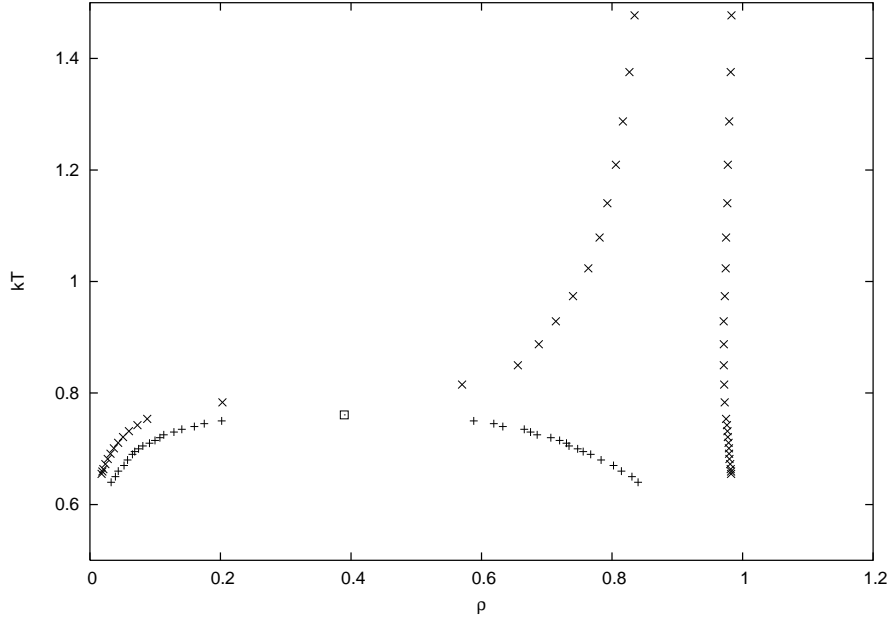}
 \caption{Phase diagram of the square well model with
 $\lambda=1.25$. Here 'x' denotes liquid-solidus line; '+'
 denotes  fluid-fluid coexistence.
 The fluid-fluid coexistence line was obtained using the Gibbs ensemble method
 with $N=600$ and $V=1500$. The liquid-solid coexistence line was obtained
 using thermodynamic integration and Gibbs-Duhem integration
 techniques.
Open square shows the position of the critical point.}
 \label{fig_CanonicalSWPhaseDiagram}
\end{figure}

In this section we summarize the algorithm of the mapping of the
canonical square well (SW) phase diagram onto the diagram of the
square well system in the presence of the solvent. The range of
particle-particle interactions and the range of the
particle-solvent interactions is considered to be the same. For
the illustration we use the SW system with $\lambda = 1.25$.
Correspondingly $\lambda_s = \lambda = 1.25$. The phase diagram of
the system without solvent is shown in the fig.
\ref{fig_CanonicalSWPhaseDiagram}. The temperature of the system
\textit{without} solvent we denote as $\tau$ in order to not to
confuse it with the temperature of the system \textit{with}
solvent, which is still denoted as $kT$.

\begin{figure}
\center
\includegraphics[width=14cm]{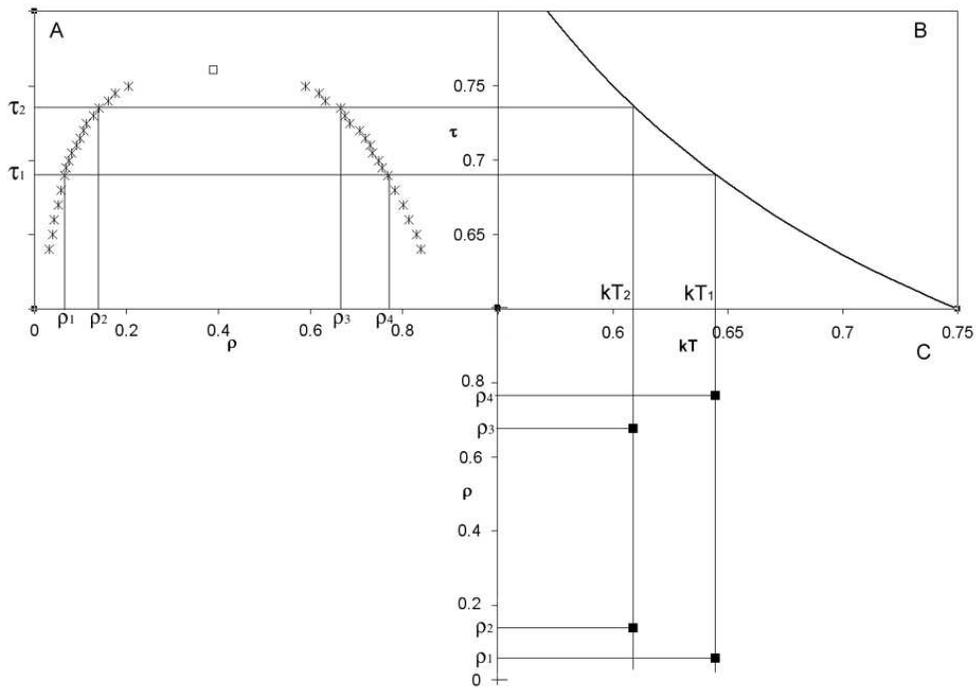}
 \caption{Mapping scheme. \textbf{A.} The fluid-fluid coexistence line
 for the square well model in the absence of solvent (same as on
 fig. \ref{fig_CanonicalSWPhaseDiagram}). \textbf{B.} The mapping curve
 for $\epsilon_w=-1$ and $\Delta s_w=-1.5$. \textbf{C.} Results of the mapping.
 Filled squares are the points on the corresponding phase diagram for the
 square well model with solvent. See text for more description. }
 \label{fig_Mapping}
\end{figure}

\begin{figure}
\center
\includegraphics[width=12cm]{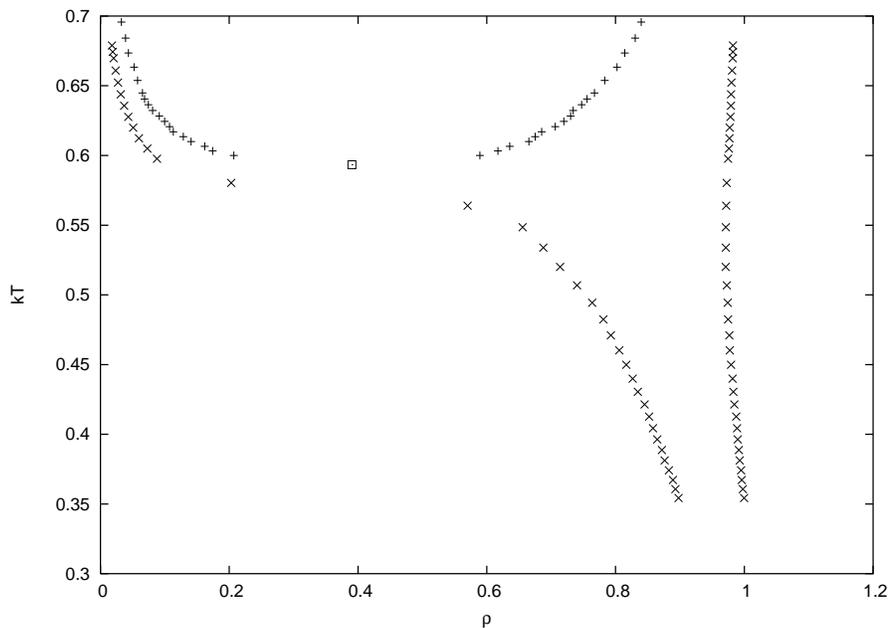}
 \caption{The phase diagram of the square well system with
 the solvent parameters $\epsilon_w = -1$ and $\Delta s_w =
 -1.5$,
 obtained by the procedure shown in  figure \ref{fig_Mapping}.
 'x' denotes liquid-solidus line; '+'
 denotes  fluid-fluid coexistence.}
 \label{fig_PhaseDiagramSolvent1}
\end{figure}

The dependence of the left hand side of (\ref{coexistence1}) on
the temperature of the system with solvent ($kT$ in equation
(\ref{coexistence1})) we will call the \textit{mapping curve}. Now
we construct the phase diagram of the system with the solvent. The
mapping curve plays an important role in this construction
process. Figure \ref{fig_Mapping} shows the mapping procedure in
detail. First we take the original phase diagram of the square
well model(fig. \ref{fig_Mapping}A). Second,  choose some
temperature $\tau_1$ and invert the mapping curve (fig.
\ref{fig_Mapping}B) to obtain the corresponding temperature of the
system with the solvent\footnote{Note that for constant
$\epsilon_w$ and $\Delta s_w$ this inversion has an analytical
solution (\ref{coexistence2}). However for temperature dependent
$\epsilon_w$ and $\Delta s_w$ as in the case of the closed loop
like phase diagram, we have to invert the mapping curve
numerically.}. Coexistent densities ($\rho_1$ and $\rho_4$ in the
fig. \ref{fig_Mapping}) corresponding to the temperature $\tau_1$
are also the coexistent densities of the system with solvent at a
temperature $kT_1$. Now choose another temperature $\tau_2$,
invert it, using the mapping curve,  to $kT_2$ and use densities
$\rho_2$ and $\rho_3$ (fig. \ref{fig_Mapping}) as the coexistent
densities at $kT_2$ for the system with solvent. By continuing
this process we obtain the  phase diagram for the system with
solvent (figure \ref{fig_PhaseDiagramSolvent1}). As one can see,
this has a lower critical point and in general an upside down
coexistence curve as  compared with fig.
\ref{fig_CanonicalSWPhaseDiagram}. The interesting feature of this
phase diagram is that at a temperature equal to
\begin{equation}
kT_{HS} = \frac{\epsilon_0 + 2\epsilon_w}{2\Delta s_w}
\label{HardSphereTemperature}
\end{equation}
the effective depth of the square well becomes zero and the
particles behave as hard spheres, with the corresponding
coexistent densities equal to the hard spheres case. In the figure
\ref{fig_PhaseDiagramSolvent1} $kT_{HS}=1/3$.

\begin{figure}
\center
\includegraphics[width=12cm]{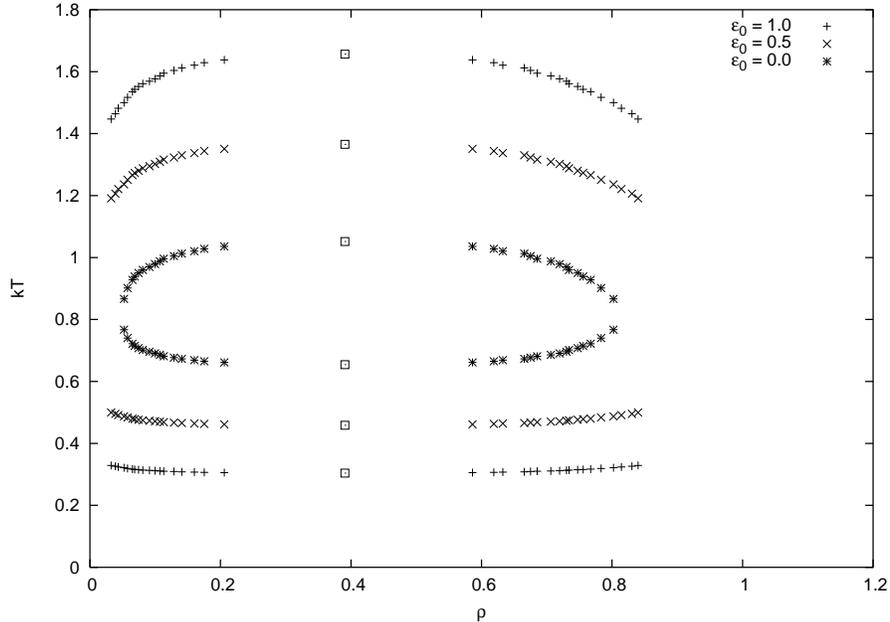}
 \caption{Fluid-fluid coexistence curves in the form of closed loop for
 the different protein-protein interaction strength $\epsilon_0$. The choice of
 solvent parameters is determined by equations (\ref{Energy_shell}) -
 (\ref{Entropy_diff}). The case with $\epsilon_0=0$ qualitatively corresponds
 to the situation described in \cite{Moelbert_03_01} where two proteins interact as hard spheres (on lattice).}
 \label{fig_closedloopsAll}
\end{figure}

The next step is to invert the mapping curves shown in the figure
\ref{fig_loopMapping}. For this purpose we have to solve
numerically equation (\ref{coexistence1}) with parameters given by
(\ref{Energy_shell}) - (\ref{Entropy_bulk}). Figure
\ref{fig_closedloopsAll} shows three fluid-fluid coexistence
curves corresponding to the three mapping curves shown on fig.
\ref{fig_loopMapping}. One can see that indeed the phase diagrams
have the form of closed loops. This is because each temperature
$\tau$ of the  system without solvent corresponds to two
temperatures $kT_1$ and $kT_2$ for the system with solvent. This
can be understood from consideration of the MLG energy level model
(fig. \ref{fig_MLG}). At low temperatures the solvent molecules
mostly occupy the lowest level, corresponding to the ordered
shell. This state corresponds to the uniform distribution of the
solute particles surrounded by the structures of solvent
molecules. At high temperatures the solvent molecules mostly
occupy the top energy level, which corresponds to the disordered
shell. So the system again tends to be in the uniform state, but
now the solvent surrounding  the solute particles is disordered.
At intermediate temperature the system tends to occupy the bulk
energy levels, which favors the phase separation. This is the
property of the MLG model that reflects the picture described in
the introduction.

\begin{figure}
\center
\includegraphics[width=12cm]{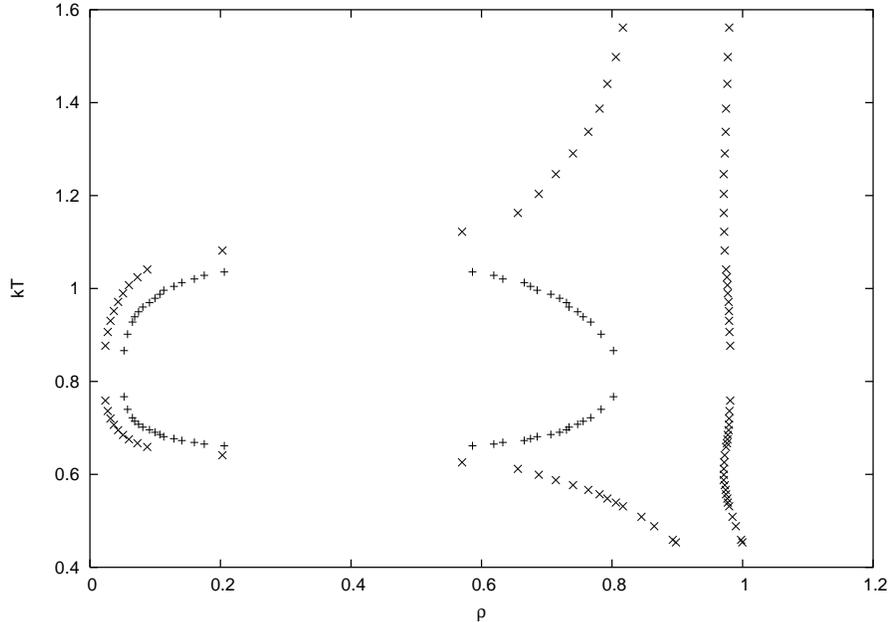}
 \caption{The fluid-fluid coexistence line as well as the
 solid-liquid coexistence line for the protein-protein interaction
 strength $\epsilon_0 = 0$ for the choice of parameters determined
 by equations (\ref{Energy_shell}) - (\ref{Entropy_diff}).}
 \label{fig_closedloopWhole}
\end{figure}

The mapping curves shown in the figure \ref{fig_loopMapping}
produce the closed loop like fluid-fluid coexistence curves.
Figure \ref{fig_closedloopWhole} shows the behavior of the
liquid-solid line in this case. The liquid-solid line also
displays a behavior similar to the fluid fluid coexistence line.
We are unaware of any experimental observations of such a
liquid-solid coexistence behavior and do not know whether it is
simply an artifact of our particular model.

\subsection{Numerical results for the MLJ system in the presence of
solvent}

    In this section consider the effect of solvent on particles interacting via
    a modified Lennard-Jones (MLJ) potential given by

  \newcolumntype{L}{>{$}1<{$}}
    \newenvironment{Cases}{\begin{array}{c}\{{1L}.}{\end{array}}
    \begin{equation}
               \large{V(r)} = \left\{ \begin{array}{cl}
                        \infty, & \mbox{$r < \sigma$}  \\
                        \frac{4\epsilon}{\alpha^{2}}(\frac{1}{[(r/\sigma)^{2}-1]^{6}}
                   -
                   \frac{\alpha}{[(r/\sigma)^{2}-1]^{3}})&\mbox{$r\ge\sigma$}.
                   \end{array} \right.
    \end{equation}
    \label{eq_MLJ}
    \noindent It has been shown that at the critical point
    the nucleation rate is many orders of magnitude
    greater than at other points in the phase diagram. This
    suggests that the nucleation of protein particles can be
    achieved near the critical point. The model has been well
    studied and its critical point accurately determined.

    For the case of the MLJ model, we again choose to use a
    solvent-solute interaction range, $\lambda_{s}$, that is
    equal to the protein-protein particle interaction range.
    Because the particles interacting via this potential have an
    effective hard-core diameter, care must be taken such that we
    choose an appropriate value for $\lambda_{s}$. We calculate
    the effective hard-core diameter using

    \begin{equation}
        \sigma_{eff} = \int_{0}^{\infty} dr[1 -
        \exp(-V_{rep}/k_{B}T)],
    \end{equation}

    \noindent where $V_{rep}$ is the repulsive part of the
    potential. The range over which the particles interact is
    controlled by the parameter $\alpha$ in \ref{eq_MLJ}. We choose $\alpha = 50$
    as in other studies.  It has been shown that for $\alpha =
    50$, one can obtain an equivalent range in terms of $\lambda$,
    the parameter denoting the range in the square well system. We
    use the value $\lambda = 1.073$. To account for the effects of
    the effective hard-core diameter, we actually use
    $\lambda_{s} = 1.26$ for the solvent-solute interaction
    range. The values $\epsilon_{w} = -1$ and $\Delta S_{w} =
    -1.5$ were used as before to obtain an upside-down phase
    diagram.

    To calculate the solid-fluid phase boundaries, we use a
    modified Gibbs-Duhem equation given by eq. 43. A coexistence
    point was calculated using free-energy methods and simulations
    of a coupled Einstein lattice. Isobaric-isothermal (NPT)
    simulations were performed in parallel for $N = 256$ particles on a
    periodic simulation cell to obtain the entire coexistence
    curve. Equilibration and production times were five million and
    ten million Monte Carlo steps, respectively.

\begin{figure}
     \rotatebox{-90}{\scalebox{.5}{\includegraphics{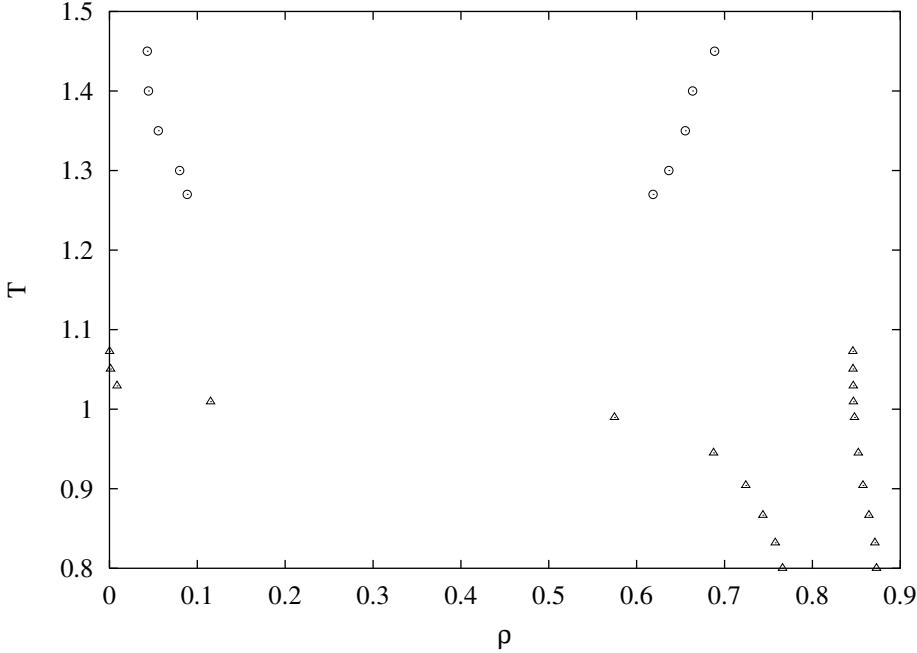}}}% Here is how to import EPS art
     \caption{\label{fig:epsart}\small {Phase diagram of the MLJ model including solvent-solute interactions. Interactions
     between the latter were mediated over the range $\lambda_{s} = 1.26$. Again 'x' denotes liquid-solidus line; '+'
     denotes  fluid-fluid coexistence.}} \label{fig_solvent}
     \end{figure}

    We calculated the fluid-fluid coexistence curve using the
    Gibbs ensemble Monte Carlo method. Two physically separated,
    but thermodynamically connected, simulation cells are allowed
    to exchange particles and undergo volume displacements such
    that the total number of particles $N = N_{1} + N_{2}$ and
    total volume $V = V_{1}+ V_{2}$ remain constant. Simulations
    were performed on an $N = 600$ particle system. Equilibration
    and production runs were fifty million and one hundred million
    Monte Carlo steps, respectively.
    Our results are shown in figure \ref{fig_solvent}.

As can be seen from the figure, the phase diagram for this model
with the particular parameters used is very similar to that for
the square well model shown in figure
\ref{fig_PhaseDiagramSolvent1}.

\section{Conclusion}

The model presented in this paper takes into account the solvent
contribution to the solute-solvent free energy of globular
proteins in solution. The contribution depends on the parameters
that describe the free energy of the solvent molecule change,
$\epsilon_w$ and $\Delta s_w$. These parameters play the role of
the solvent enthalpy and entropy change per solvent particle as
the molecule goes from the bulk to the vicinity of the protein
molecule. The solvent enthalpy and entropy change per solute
particle upon changing from the fluid phase to the solid
phase(parameters $\Delta H_{solvent}$ and $\Delta S_{solvent}$
from (\ref{EnergyChangeSeparation})) can be calculated as equal to
$\epsilon_w$ and $\Delta s_w$ times the difference of the average
number of contacts per molecule in these two phases.
\begin{eqnarray}
\Delta H_{solvent} = \epsilon_w( <n_c^{fluid}> - <n_c^{solute}>) \notag \\
\Delta S_{solvent} = \Delta s_w( <n_c^{fluid}> - <n_c^{solute}>)
\notag
\end{eqnarray}

These relationships allow us to relate $\epsilon_w$ and $\Delta
s_w$ to the three cases of the solubility dependence described in
the introduction - normal solubility dependence, retrograde
solubility dependence and  constant solubility (as in the case of
apoferritin). All we need is the sign of the total enthalpy change
$\Delta H$. For the square well case, $\epsilon_w > -\epsilon_0/2$
is a condition for $\Delta H$ being negative and therefore the
solubility curve being normal. If the range of attraction is short
this qualitatively describes the lysozyme phase diagram. The
condition $\epsilon_w < -\epsilon_0/2$ is a condition for $\Delta
H$ being positive and therefore the solubility curve being
retrograde. This corresponds to the case of the HbC solubility
curve (fig. \ref{fig_HbCSolubility}). Indeed we can see that the
liquidus line in the figure \ref{fig_PhaseDiagramSolvent1} has
qualitatively the same behavior as the HbC solubility curve. If
$\epsilon_w \cong\epsilon/2$, so that the enthalpy change is
small, then the solubility curve is almost vertical, with the sign
of the slope determined by the sign of the enthalpy change.

The advantage of this simplified model is that in the particular
case of the square well, with the particle-particle range of
interaction equal to the width of the shell region around the
particle,  one can obtain the phase diagram by a mapping of the
square well phase diagram without solvent. Therefore one doesn't
have to perform Monte Carlo simulations or theoretical
approximations to obtain the phase diagram for the model with
solvent.

The exact mapping is also possible for the hard sphere model with
solvent. In this case $\epsilon_0 = 0$. The hard sphere case is
interesting in two aspects. First, one can obtain an exact mapping
procedure for any solvent-solute interaction. Second, the hard
sphere interaction doesn't include any initial solvent
contribution (unlike the case with an effective solute-solute
interaction). This can be the case of noninteracting neutral
colloidal particles in the solvent. If we put such particles into
water, the hydrogen bonds break and rearrange,  as explained in
the introduction and in \cite{Moelbert_03_01}.  Therefore we can
use the MLG model to get solvent parameters $\epsilon_w$ and
$\Delta s_w$. We can consider the hard sphere interaction to be a
square well with $\epsilon_0=0$ and $\lambda=\lambda_s$. Then
using a mapping curve as in the figure \ref{fig_loopMapping} (the
one with $\epsilon_0=0$) we can obtain the phase diagram of the
neutral, noninteracting spherical particles in water. If the
critical temperature of the square well system with a range of
interaction equal to $\lambda_s$ intersects the mapping curve
(fig. \ref{fig_loopMapping}), the colloidal system has a closed
loop type phase diagram (fig \ref{fig_closedloopsAll} with
$\epsilon_0=0$). As the value of $\lambda_s$ decreases (the size
of the particle increases), the fluid-fluid coexistence curve
shrinks and disappears. So larger noninteracting particles don't
have a fluid-fluid phase separation in this model. However, if the
particles starts to interact, the mapping curve lowers (fig
\ref{fig_loopMapping}) and phase separation may occur again.

For the case of constant solvent parameters we can have either an
upper critical point with a normal solubility dependence, or a
lower critical point with a retrograde solubility dependence. The
temperature dependent parameters allow us to have a combination of
these behaviors. A natural way to derive the temperature dependent
solvent parameters for water is to use  the MLG model. This model,
together with our solvent-solute interaction, has closed loop
phase diagrams, similar to \cite{Moelbert_03_01}.

\section{Acknowledgements}
This work was supported by NSF grant DMR-0302598. One of us (A.S.)
wishes to acknowledge the support and hospitality of the
Theoretical Division of the Los Alamos National Laboratory, and a
second (D.S.R.) wishes to acknowledge the support of the German
Fulbright Commission and to thank the Helmholtz Institute for
Radiation and Nuclear Physics of the University of Bonn, for the
use of their computing facilities.

\end{document}